%% file: paper.tex
\documentclass[aps,prl,showpacs,twocolumn,groupedaddress]{revtex4}  
\usepackage{graphicx}  
\usepackage{dcolumn}   
\usepackage{bm}        
\usepackage{amssymb}   

\usepackage{epsfig}
\usepackage{cancel}

\include{comm}

\begin{document}


\hspace{5.2in} \mbox{Fermilab-Pub-05-075-E}

\title{Search for Supersymmetry via Associated Production of Charginos
and Neutralinos in Final States with Three Leptons}
\input list_of_authors_r2.tex  
\date{April 18, 2005}

\begin{abstract}
\input{abstract}
\end{abstract}

\maketitle

\input{main}

\end{document}

%% file: comm.tex
\newcommand{\ppbar}{\ensuremath{p\bar{p}}}
\newcommand{\tanb}{\ensuremath{\tan\beta}}
\newcommand{\cha}{\ensuremath{\tilde{\chi}^{\pm}}}
\newcommand{\chiz}{\ensuremath{\tilde{\chi}^0}}

\newcommand{\mcha}{\ensuremath{m_{\tilde{\chi}^{\pm}_1}}}
\newcommand{\mchione}{\ensuremath{m_{\tilde{\chi}^0_1}}}
\newcommand{\mchitwo}{\ensuremath{m_{\tilde{\chi}^0_2}}}
\newcommand{\mslep}{\ensuremath{m_{\tilde{\ell}}}}

\newcommand{\met}{\mbox{\ensuremath{\not \hspace{-1.0mm} E_T}}}

\newcommand{\dr}{\ensuremath{\Delta{\cal R}}}
\newcommand{\pbinv}{\ensuremath{\mathrm{pb^{-1}}}}

\newcommand{\ptone}{\ensuremath{p_T^{\ell1}}}
\newcommand{\pttwo}{\ensuremath{p_T^{\ell2}}}
\newcommand{\ptthree}{\ensuremath{p_T^{\ell3}}}
\newcommand{\mll}{\ensuremath{m_{\ell^{1\!,2}\ell^3}}}

\newcommand{\sigbr}{\ensuremath{\sigma\times\mathrm{BR}(3\ell)}}

\newcommand{\eel}{\ensuremath{ee\ell}}
\newcommand{\mumul}{\ensuremath{\mu\mu \ell}}
\newcommand{\emul}{\ensuremath{e\mu \ell}}
\newcommand{\lsmumu}{\ensuremath{\mu^{\pm}\mu^{\pm}}}

%% file: list_of_authors_r2.tex
%
\author{                                                                      
V.M.~Abazov,$^{35}$                                                           
B.~Abbott,$^{72}$                                                             
M.~Abolins,$^{63}$                                                            
B.S.~Acharya,$^{29}$                                                          
M.~Adams,$^{50}$                                                              
T.~Adams,$^{48}$                                                              
M.~Agelou,$^{18}$                                                             
J.-L.~Agram,$^{19}$                                                           
S.H.~Ahn,$^{31}$                                                              
M.~Ahsan,$^{57}$
% ...
\\                                                                            
\vskip 0.30cm                                                                 
\centerline{(D\O\ Collaboration)}                                             
\vskip 0.30cm                                                                 
}                                                                             
\affiliation{                                                                 
\centerline{$^{1}$Universidad de Buenos Aires, Buenos Aires, Argentina}       
\centerline{$^{2}$LAFEX, Centro Brasileiro de Pesquisas F{\'\i}sicas,         
                  Rio de Janeiro, Brazil}                                     
\centerline{$^{3}$Universidade do Estado do Rio de Janeiro,                   
                  Rio de Janeiro, Brazil}                                     
\centerline{$^{4}$Instituto de F\'{\i}sica Te\'orica, Universidade            
                  Estadual Paulista, S\~ao Paulo, Brazil}                     
\centerline{$^{5}$University of Alberta, Edmonton, Alberta, Canada,           
               Simon Fraser University, Burnaby, British Columbia, Canada,}   
\centerline{York University, Toronto, Ontario, Canada, and                    
         McGill University, Montreal, Quebec, Canada}                         
\centerline{$^{6}$Institute of High Energy Physics, Beijing,                  
                  People's Republic of China}                                 
\centerline{$^{7}$University of Science and Technology of China, Hefei,       
                  People's Republic of China}                                 
\centerline{$^{8}$Universidad de los Andes, Bogot\'{a}, Colombia}             
\centerline{$^{9}$Center for Particle Physics, Charles University,            
                  Prague, Czech Republic}                                     
\centerline{$^{10}$Czech Technical University, Prague, Czech Republic}        
\centerline{$^{11}$Institute of Physics, Academy of Sciences, Center          
                  for Particle Physics, Prague, Czech Republic}               
\centerline{$^{12}$Universidad San Francisco de Quito, Quito, Ecuador}        
\centerline{$^{13}$Laboratoire de Physique Corpusculaire, IN2P3-CNRS,         
                 Universit\'e Blaise Pascal, Clermont-Ferrand, France}        
\centerline{$^{14}$Laboratoire de Physique Subatomique et de Cosmologie,      
                  IN2P3-CNRS, Universite de Grenoble 1, Grenoble, France}     
\centerline{$^{15}$CPPM, IN2P3-CNRS, Universit\'e de la M\'editerran\'ee,     
                  Marseille, France}                                          
\centerline{$^{16}$Laboratoire de l'Acc\'el\'erateur Lin\'eaire,              
                  IN2P3-CNRS, Orsay, France}                                  
\centerline{$^{17}$LPNHE, IN2P3-CNRS, Universit\'es Paris VI and VII,         
                  Paris, France}                                              
\centerline{$^{18}$DAPNIA/Service de Physique des Particules, CEA, Saclay,    
                  France}                                                     
\centerline{$^{19}$IReS, IN2P3-CNRS, Universit\'e Louis Pasteur, Strasbourg,  
                France, and Universit\'e de Haute Alsace, Mulhouse, France}   
\centerline{$^{20}$Institut de Physique Nucl\'eaire de Lyon, IN2P3-CNRS,      
                   Universit\'e Claude Bernard, Villeurbanne, France}         
\centerline{$^{21}$III. Physikalisches Institut A, RWTH Aachen,               
                   Aachen, Germany}                                           
\centerline{$^{22}$Physikalisches Institut, Universit{\"a}t Bonn,             
                  Bonn, Germany}                                              
\centerline{$^{23}$Physikalisches Institut, Universit{\"a}t Freiburg,         
                  Freiburg, Germany}                                          
\centerline{$^{24}$Institut f{\"u}r Physik, Universit{\"a}t Mainz,            
                  Mainz, Germany}                                             
\centerline{$^{25}$Ludwig-Maximilians-Universit{\"a}t M{\"u}nchen,            
                   M{\"u}nchen, Germany}                                      
\centerline{$^{26}$Fachbereich Physik, University of Wuppertal,               
                   Wuppertal, Germany}                                        
\centerline{$^{27}$Panjab University, Chandigarh, India}                      
\centerline{$^{28}$Delhi University, Delhi, India}                            
\centerline{$^{29}$Tata Institute of Fundamental Research, Mumbai, India}     
\centerline{$^{30}$University College Dublin, Dublin, Ireland}                
\centerline{$^{31}$Korea Detector Laboratory, Korea University,               
                   Seoul, Korea}                                              
\centerline{$^{32}$CINVESTAV, Mexico City, Mexico}                            
\centerline{$^{33}$FOM-Institute NIKHEF and University of                     
                  Amsterdam/NIKHEF, Amsterdam, The Netherlands}               
\centerline{$^{34}$Radboud University Nijmegen/NIKHEF, Nijmegen, The          
                  Netherlands}                                                
\centerline{$^{35}$Joint Institute for Nuclear Research, Dubna, Russia}       
\centerline{$^{36}$Institute for Theoretical and Experimental Physics,        
                  Moscow, Russia}                                             
\centerline{$^{37}$Moscow State University, Moscow, Russia}                   
\centerline{$^{38}$Institute for High Energy Physics, Protvino, Russia}       
\centerline{$^{39}$Petersburg Nuclear Physics Institute,                      
                   St. Petersburg, Russia}                                    
\centerline{$^{40}$Lund University, Lund, Sweden, Royal Institute of          
                   Technology and Stockholm University, Stockholm,            
                   Sweden, and}                                               
\centerline{Uppsala University, Uppsala, Sweden}                              
\centerline{$^{41}$Lancaster University, Lancaster, United Kingdom}           
\centerline{$^{42}$Imperial College, London, United Kingdom}                  
\centerline{$^{43}$University of Manchester, Manchester, United Kingdom}      
\centerline{$^{44}$University of Arizona, Tucson, Arizona 85721, USA}         
\centerline{$^{45}$Lawrence Berkeley National Laboratory and University of    
                  California, Berkeley, California 94720, USA}                
\centerline{$^{46}$California State University, Fresno, California 93740, USA}
\centerline{$^{47}$University of California, Riverside, California 92521, USA}
\centerline{$^{48}$Florida State University, Tallahassee, Florida 32306, USA} 
\centerline{$^{49}$Fermi National Accelerator Laboratory, Batavia,            
                   Illinois 60510, USA}                                       
\centerline{$^{50}$University of Illinois at Chicago, Chicago,                
                   Illinois 60607, USA}                                       
\centerline{$^{51}$Northern Illinois University, DeKalb, Illinois 60115, USA} 
\centerline{$^{52}$Northwestern University, Evanston, Illinois 60208, USA}    
\centerline{$^{53}$Indiana University, Bloomington, Indiana 47405, USA}       
\centerline{$^{54}$University of Notre Dame, Notre Dame, Indiana 46556, USA}  
\centerline{$^{55}$Iowa State University, Ames, Iowa 50011, USA}              
\centerline{$^{56}$University of Kansas, Lawrence, Kansas 66045, USA}         
\centerline{$^{57}$Kansas State University, Manhattan, Kansas 66506, USA}     
\centerline{$^{58}$Louisiana Tech University, Ruston, Louisiana 71272, USA}   
\centerline{$^{59}$University of Maryland, College Park, Maryland 20742, USA} 
\centerline{$^{60}$Boston University, Boston, Massachusetts 02215, USA}       
\centerline{$^{61}$Northeastern University, Boston, Massachusetts 02115, USA} 
\centerline{$^{62}$University of Michigan, Ann Arbor, Michigan 48109, USA}    
\centerline{$^{63}$Michigan State University, East Lansing, Michigan 48824,   
                   USA}                                                       
\centerline{$^{64}$University of Mississippi, University, Mississippi 38677,  
                   USA}                                                       
\centerline{$^{65}$University of Nebraska, Lincoln, Nebraska 68588, USA}      
\centerline{$^{66}$Princeton University, Princeton, New Jersey 08544, USA}    
\centerline{$^{67}$Columbia University, New York, New York 10027, USA}        
\centerline{$^{68}$University of Rochester, Rochester, New York 14627, USA}   
\centerline{$^{69}$State University of New York, Stony Brook,                 
                   New York 11794, USA}                                       
\centerline{$^{70}$Brookhaven National Laboratory, Upton, New York 11973, USA}
\centerline{$^{71}$Langston University, Langston, Oklahoma 73050, USA}        
\centerline{$^{72}$University of Oklahoma, Norman, Oklahoma 73019, USA}       
\centerline{$^{73}$Brown University, Providence, Rhode Island 02912, USA}     
\centerline{$^{74}$University of Texas, Arlington, Texas 76019, USA}          
\centerline{$^{75}$Southern Methodist University, Dallas, Texas 75275, USA}   
\centerline{$^{76}$Rice University, Houston, Texas 77005, USA}                
\centerline{$^{77}$University of Virginia, Charlottesville, Virginia 22901,   
                   USA}                                                       
\centerline{$^{78}$University of Washington, Seattle, Washington 98195, USA}  
}                                                                             

%% file: abstract.tex
A search for associated production of charginos and neutralinos is
performed using data recorded with the D\O\ detector at a
\ppbar\ center-of-mass energy of 1.96~TeV at the Fermilab Tevatron
Collider. This analysis considers final states with missing transverse
energy and three leptons, of which at least two are electrons or muons.
No evidence for supersymmetry is found in a dataset
corresponding to an integrated luminosity of 320~\pbinv. 
Limits on the product of the production cross section and leptonic
branching fraction are set. For the minimal supergravity model,
a chargino lower mass limit of 117~GeV at the 95\% C.L. is derived in
regions of parameter space with enhanced leptonic branching
fractions. 

%% file: main.tex
Supersymmetry (SUSY) predicts the existence of a new particle for each of the
standard model particles, differing by half a unit in spin but
otherwise sharing the same quantum numbers.
No supersymmetric particles have been observed so far, and it is
therefore generally assumed that they are heavier than their standard
model partners. Experiments at the CERN LEP Collider have set lower limits
on the masses of SUSY particles, excluding in particular charginos
with masses below 103.5~GeV as well as sleptons
with masses below about 95~GeV~\cite{LEP_limits} in the framework of
the minimal supersymmetric model.
Due to its high center-of-mass energy of 1.96~TeV, the
Tevatron \ppbar\ collider may produce SUSY particles with masses above
these limits. 
A search for SUSY can be performed via the associated
production of charginos and neutralinos.
The lightest chargino~$\cha_1$ and the second-lightest neutralino~$\chiz_2$
are assumed to decay via exchange of vector bosons or sleptons into the
lightest neutralino and standard model fermions. Assuming conservation
of $R$-parity, the lightest neutralino is stable and 
can only be detected indirectly.

This Letter reports on a search for $\ppbar\to\cha_1\chiz_2$
in final states with missing transverse energy and three charged
leptons ($e$, $\mu$ or $\tau$), of which at least two are electrons or
muons. The analysis is  
based on a dataset recorded with the D\O\ detector between
March 2002 and July 2004, corresponding to an integrated luminosity of
320~\pbinv. Previous searches in 
this channel have been performed by the CDF and D\O\ collaborations
with Tevatron Run~I data~\cite{run1_limits}.

The D\O\ detector consists of a central tracking system surrounded by
a liquid-argon sampling calorimeter and a system of muon
detectors~\cite{run2det}. Charged particles are reconstructed using multiple layers
of silicon detectors as well as eight double layers of scintillating
fibers in the 2~T magnetic field of a superconducting
solenoid. The D\O\ calorimeter provides hermetic coverage up to
pseudorapidities~$|\eta|\approx 4$ in a semi-projective tower geometry
with longitudinal segmentation. 
After passing through the calorimeter, muons are
detected in three layers of tracking detectors and scintillation
counters.

Events containing electrons or muons are selected for offline analysis
by a real-time three-stage trigger system. A set of single and
dilepton triggers has been used to tag the presence of electrons and
muons based on their characteristic energy deposits in the calorimeter,
the presence of high-momentum tracks in the tracking system, and
hits in the muon detectors. 

SUSY and standard model processes are modeled using the
{\sc pythia}~\cite{pythia} Monte Carlo (MC) generator and a detailed
simulation of the 
detector geometry and response based on {\sc geant}~\cite{geant}.
Multiple interactions per crossing as well as pile-up of signals in
the calorimeter have been
simulated. The MC events are then processed using the same
reconstruction and analysis programs that are used for the data.
The background predictions are normalized using cross-section
calculations at next-to-leading order (NLO) and next-to-NLO (for
Drell-Yan production) with CTEQ6.1M~\cite{cteq6} parton distribution
functions (PDFs).  

Background from multijet production is estimated from data.
For this, samples dominated by multijet background have been defined
that are identical to the search sample except for reversed lepton
identification requirements.
These samples are normalized 
at an early stage of the selection in a region of phase space
dominated by multijet production.

Selection criteria are optimized to obtain the best average expected
limit assuming that no signal will be observed. 
Limits are calculated at the 95\% C.L. using the modified
frequentist approach~\cite{cls}.
The optimization of selection cuts is based on signals
inspired by minimal supergravity (mSUGRA) \cite{SUGRA} with
$\cha_1$, $\chiz_2$ and slepton masses in the range 110--130~GeV. Due
to the large production cross section and leptonic branching
fraction via slepton exchange, this mass range is of particular interest
for a search in the trilepton channel. 
In the following discussion of the selection, as a representative example,
a signal is used with common scalar mass $m_0=84$~GeV, common fermion
mass $m_{1/2}=176$~GeV, ratio of Higgs vacuum expectation values $\tanb=3$,
Higgs mass parameter $\mu>0$, and no slepton mixing,
which corresponds to a chargino mass of 110~GeV and $\sigbr=0.265$~pb.

Four different selections are defined depending on the lepton
content of the final state: two electrons plus lepton (\eel\ selection); two
muons plus lepton (\mumul); two muons of the same charge (\lsmumu);
and one electron, one muon plus lepton (\emul).
The selection criteria are summarized in Table~\ref{tab:cuts} and are 
discussed in more detail below.
\begin{table}
\caption{\label{tab:cuts}Selection criteria for the four analyses
(all energies and momenta in GeV, angles in radians), see text
for further details.} 
\begin{ruledtabular}
\renewcommand{\arraystretch}{1.2}
\begin{tabular}{cccccc}
 & Selection Cut & \eel 
              & \mumul 
              & \lsmumu
              & \emul\\
\hline
I & \ptone, \pttwo & 
  $>$12, $>$8 & $>$11, $>$5 & $>$11, $>$5 & $>$12, $>$8\footnotemark[1]\\
\hline
 & $m_{\ell\ell}$ &
 $\in$[18,60] & $\in$[15,50] & $<$80 & --\\
II & $\Delta\phi_{\ell\ell}$ &
 $<$2.9 & -- & $<$2.7 & --\\
\hline
 & \met\ & 
 $>$22 & $>$22 & $>$22 & -- \\
 & Sig(\met) & 
 $>$8 & $>$8 & $>$8 & $>$15 \\
 & $m_T^{\mathrm{min}}$ &
 $>$20 & $>$20 & -- & $\in$[25,90]\\
III & jet-veto &
 $H_T$$<$80 & -- & -- & $H_T$$<$40\\
\hline 
IV & \ptthree &
 $>$4 & $>$3 & -- & $>$7\\
\hline
 & \mll &
 -- & $<$70 & $\notin$[70,110]\footnotemark[2] & $\notin$[60,120]\\ 
 & $|\Sigma_{p_T}|/\ptthree$ & -- & $\in$[0.3,3.0] & -- & --\\
V & \met\ $\times$ \ptthree & 
 $>$220 & $>$150 & $>$300\footnotemark[3] & --\\
\end{tabular}
\renewcommand{\arraystretch}{1.0}
\end{ruledtabular}
\footnotetext[1]{\ptone\ and \pttwo\ are electron and muon $p_T$, respectively.}
\footnotetext[2]{Opposite-sign muons only.}
\footnotetext[3]{Using \pttwo\ instead of \ptthree.}
\end{table}

Isolated electrons are identified based on their characteristic energy
deposition in the calorimeter, including the fraction of energy
deposited in the electromagnetic portion of the calorimeter and their
transverse shower profile inside a cone of radius $\dr
=\sqrt{(\Delta\phi)^2+(\Delta\eta)^2}<0.4$ around the direction of the
electron.
In addition it is required that a track points to the energy
deposition in the calorimeter and that its momentum and the
calorimeter energy are consistent with the same electron
energy. 
Remaining backgrounds from jets and photon conversions
are suppressed based on the track activity within $\dr=0.4$ around
the track direction and by requiring the track associated with electron
candidates to have associated hits in the innermost layers of the silicon
detector.

\begin{figure}
\epsfig{file=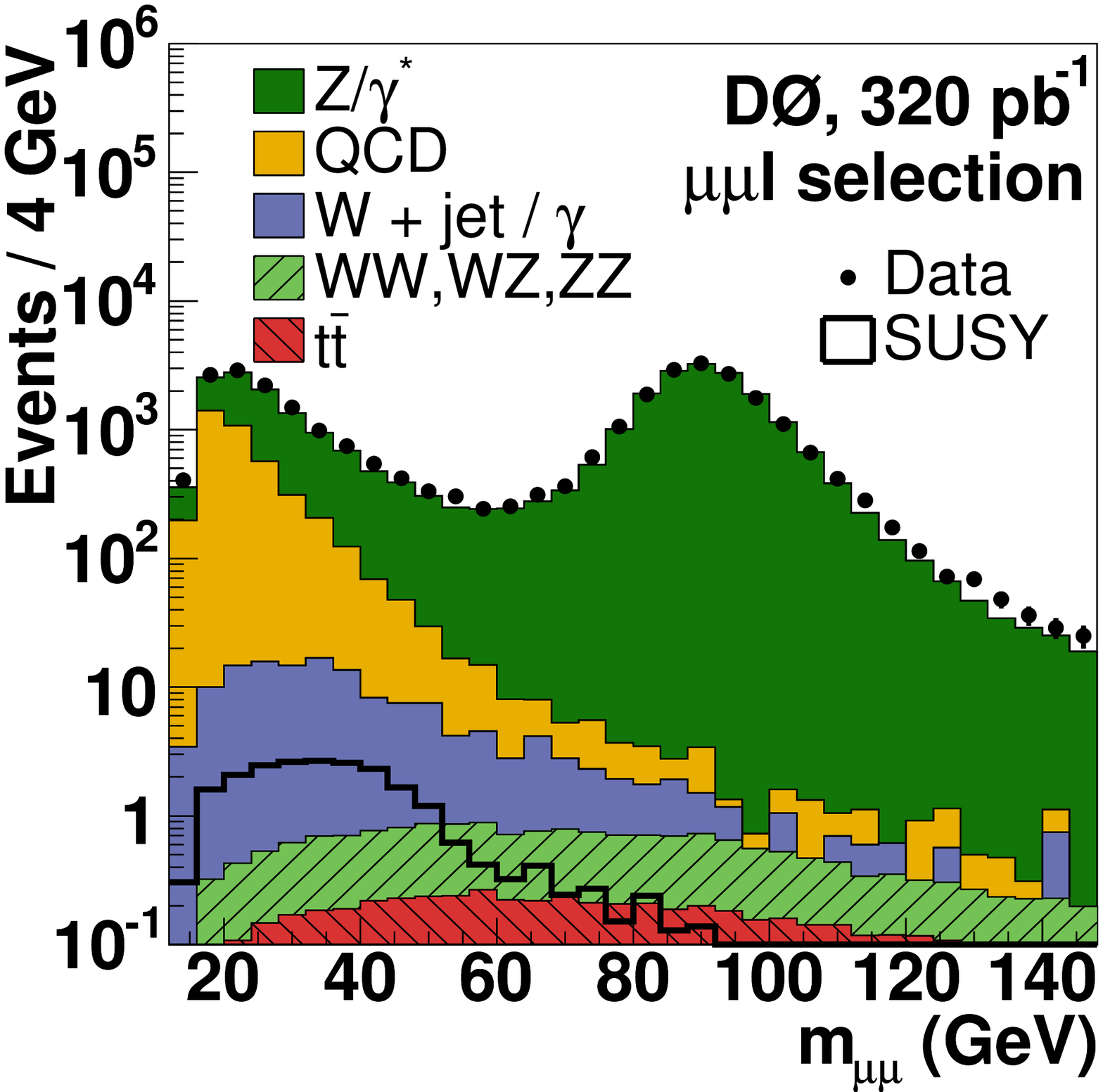,width=0.5\columnwidth,clip=true} \hspace{-0.2cm}
\epsfig{file=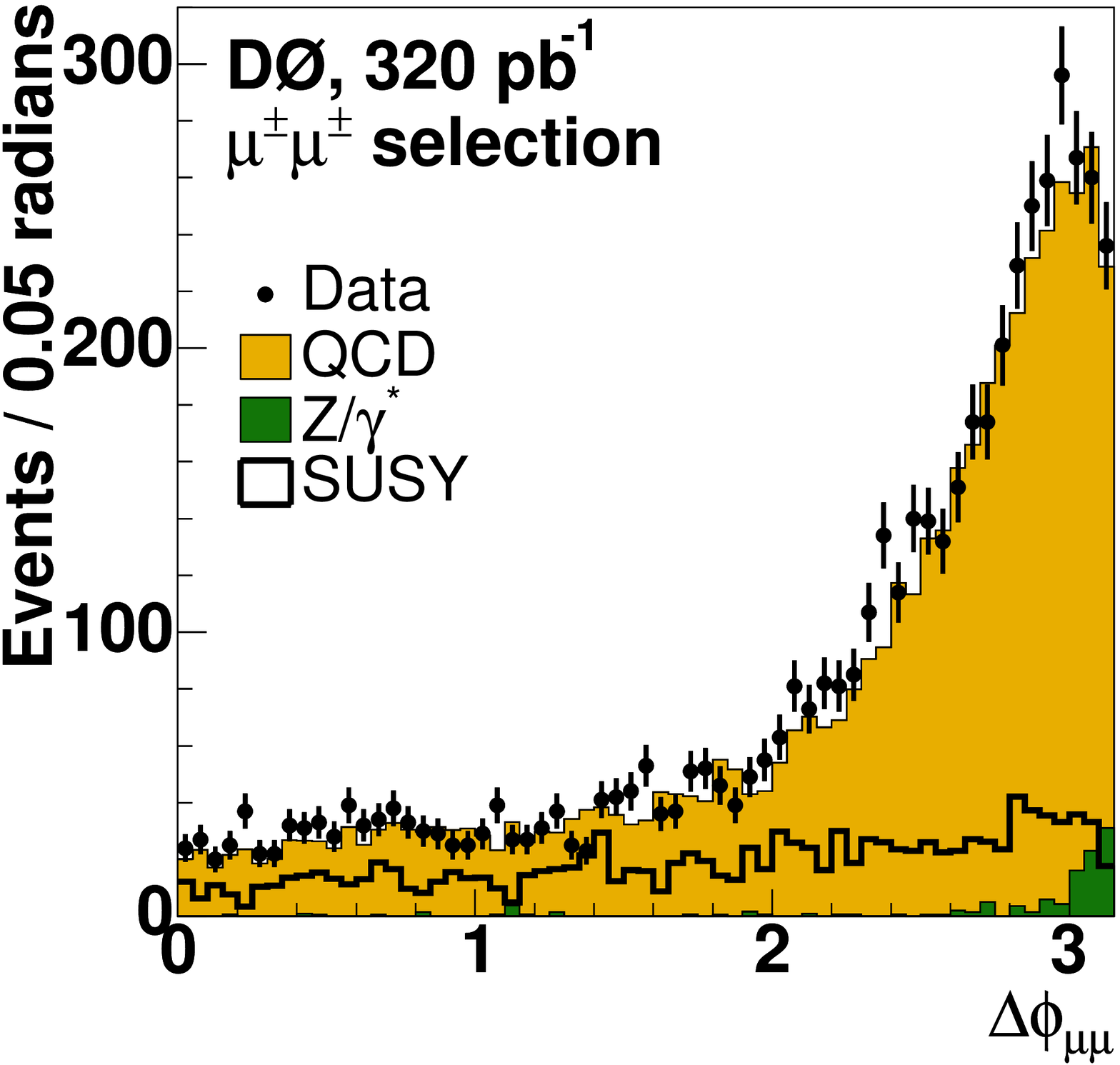,width=0.5\columnwidth,clip=true}
\caption{\label{fig:selection1}Invariant di-muon mass $m_{\mu\mu}$ (left,
\mumul\ selection) and azimuthal di-muon opening angle~$\Delta\phi_{\mu\mu}$
(right, \lsmumu\ selection) for data (points), standard 
model backgrounds (shaded histograms), and SUSY signal (open histogram).}
\end{figure}
Muons are reconstructed by finding tracks that 
point to patterns of hits in the muon system. 
Non-isolated muons from backgrounds with heavy-flavor jets are
rejected by requiring the sum of track $p_T$ inside a cone with
$\dr=0.5$ around 
the muon direction to be less than 4~GeV ({\em loose muons}) or less
than 2.5~GeV ({\em tight muons}). Tight muons are also required
to have less than 2.5~GeV deposited in the calorimeter in a
hollow cone $0.1<\dr<0.4$ around the muon direction.

Electron and muon reconstruction efficiencies have been measured using
leptonic $Z$ boson decays collected by single-lepton triggers.
The electron and muon trigger
efficiencies have been measured in data and translate to an event
trigger efficiency close to 100\% (85\%) for signal events passing
offline analysis requirements in the \eel, \emul\ (\lsmumu, \mumul)
selections. 

Each selection requires two identified leptons with minimum transverse
momenta \ptone\ and \pttwo, using one loose muon for the \emul, one
loose and one tight muon for the \mumul\ and two tight muons for the
\lsmumu\ selection.
Further selection cuts exploiting the difference in event kinematics 
and topology are applied as summarized in Table~\ref{tab:cuts}.
Di-electron and di-muon backgrounds from Drell-Yan and $Z$ boson production as
well as multijet background are suppressed using 
cuts on the invariant dilepton mass~$m_{\ell\ell}$ as well as the azimuthal opening
angle~$\Delta\phi_{\ell\ell}$. As illustrated in Fig.~\ref{fig:selection1},
a large fraction of these events can be 
rejected by removing events containing leptons that are back-to-back
in azimuthal angle as well as 
events with $m_{\ell\ell}$ close to the $Z$ boson mass.

\begin{figure}
\epsfig{file=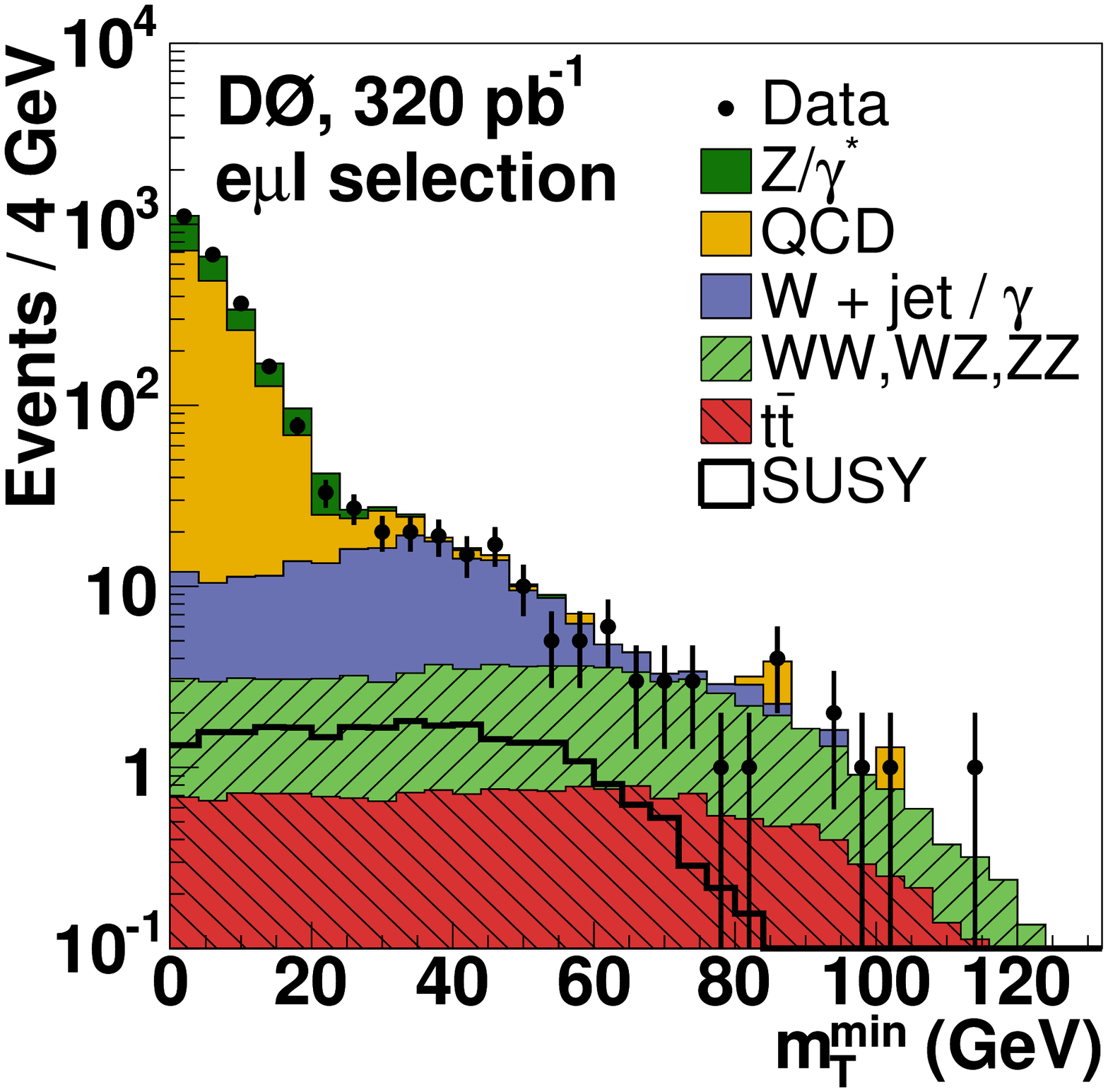,width=0.5\columnwidth,clip=true} \hspace{-0.2cm}
\epsfig{file=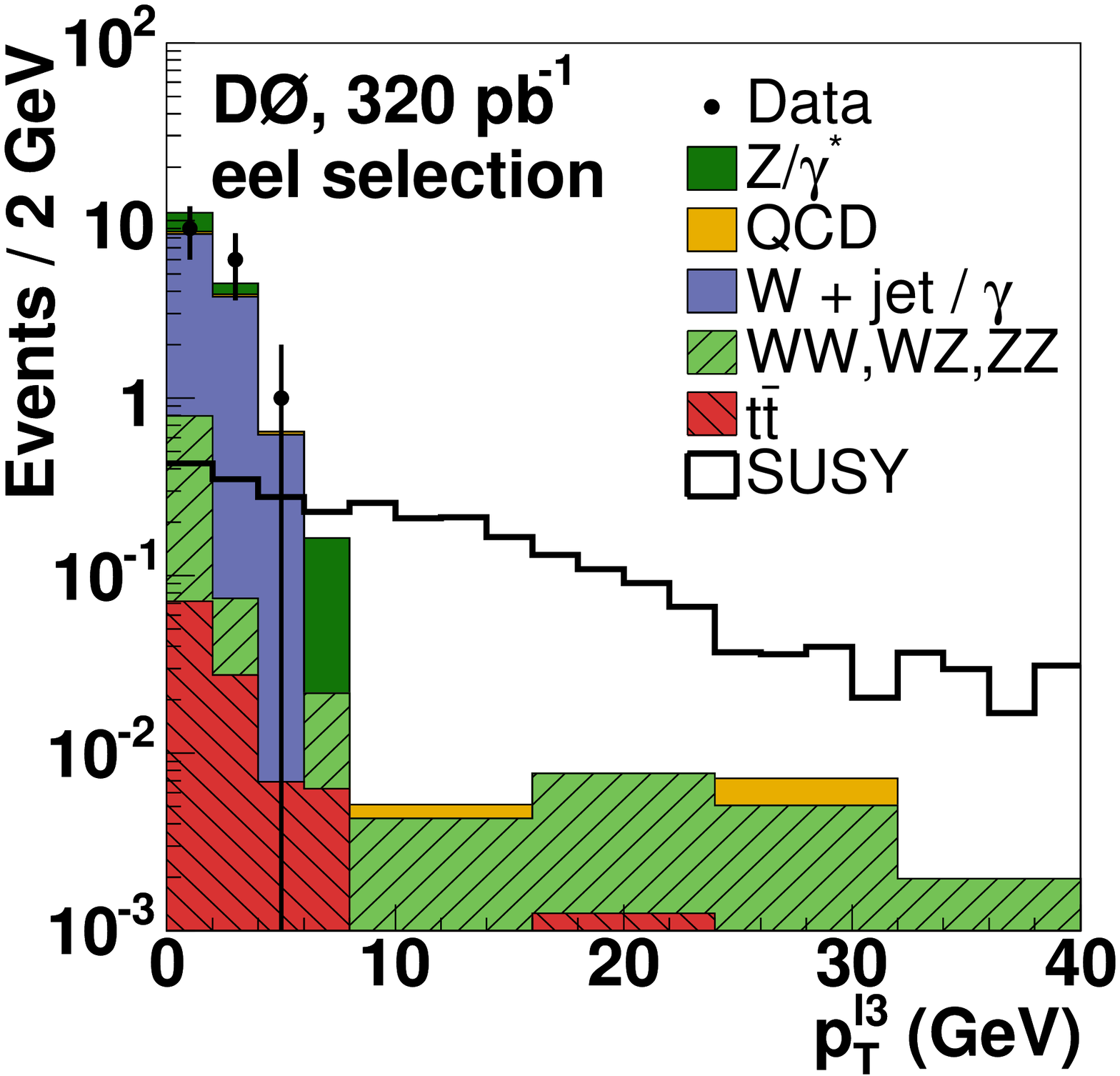,width=0.5\columnwidth,clip=true}
\caption{\label{fig:selection2}Minimum transverse
mass $m_T^{min}$ (left, \emul\ selection) and transverse momentum of
the third track
$p_T^{\ell3}$ (right, \eel\ selection) for data 
(points), standard model backgrounds (shaded histograms), and SUSY
signal (open histogram).} 
\end{figure}

A further reduction in dilepton and multijet backgrounds can be
achieved by requiring missing transverse energy~\met\ in an event. 
This is calculated as the vectorial sum of energy depositions in
calorimeter cells and then adjusted using energy response
corrections for reconstructed electrons, muons and jets.
Jets are defined using an iterative seed-based cone algorithm,
clustering calorimeter energy within 
$\dr=0.5$. The jet energy calibration has been determined from
transverse momentum balance in photon plus jet events.
For background events, an imbalance in transverse energy can
be generated by mismeasurements of jet or lepton energies. Therefore,
events in which the \met\ direction is aligned with the lepton are
removed using a cut on the minimum transverse mass
$m_T^{\mathrm{min}}=\mathrm{min}(m_T^{\ell1,\not\hspace{-0.02mm}
E_T},m_T^{\ell2,\not\hspace{-0.02mm} E_T})$ as shown in  Fig.~\ref{fig:selection2}.
In addition, events are rejected if they contain jets with transverse
energies above 15~GeV and have a small significance Sig(\met), which is defined
by normalizing the \met\ to $\sigma(E_T^j\!\parallel\!\met)$, a
measure of the jet energy resolution projected  
onto the \met\ direction:
\begin{eqnarray}
\mathrm{Sig}(\met) = \frac{\met}{\sqrt{\sum_{\mathrm{jets}}
\sigma^2_{E_T^j \parallel \,\met}}}
\label{eq:sigmet}.\nonumber
\end{eqnarray}
Most of the remaining background from $t\bar{t}$ production 
can be rejected by removing events with large $H_T$,
defined as the scalar sum of the transverse energies of all jets with
$E_T>15$~GeV. 

\begin{table*}
\caption{\label{tab:numbers}Number of events observed in data and
expected for background and reference signal (see text) at various
stages of the selection, 
with statistical and systematic uncertainties added in quadrature. Each
row corresponds to a group of cuts, as detailed in
Table~\ref{tab:cuts}. 
}
\begin{ruledtabular}
\begin{tabular}{crrrrrrrrrrrr}
Cut & \multicolumn{3}{c}{\eel\ selection}
              & \multicolumn{3}{c}{\mumul\ selection}
              & \multicolumn{3}{c}{\lsmumu\ selection}
              & \multicolumn{3}{c}{\emul\ selection}\\
 & Data & Backgrd. & Signal
 & Data & Backgrd. & Signal
 & Data & Backgrd. & Signal
 & Data & Backgrd. & Signal\\
\hline
I & 33468 & 32000$\pm$3500 & 8.8$\pm$0.8
  & 40489 & 40400$\pm$3100 & 7.7$\pm$0.9
  & 201   & 235$\pm$22     & 1.7$\pm$0.2
  & 2588  & 2600$\pm$290   & 8.9$\pm$0.7\\
II & 3921  & 3990$\pm$470  & 6.2$\pm$0.6
   & 12520 & 11750$\pm$710 & 6.1$\pm$0.7
   & 125   & 110$\pm$12    & 1.3$\pm$0.1
   & 2588  & 2600$\pm$290  & 8.9$\pm$0.7\\
III & 46  & 45$\pm$12   & 4.0$\pm$0.4
    & 135 & 182$\pm$38  & 3.3$\pm$0.4
    & 7   & 5.7$\pm$1.6 & 0.93$\pm$0.17
    & 95  & 95$\pm$11   & 4.2$\pm$0.3\\
IV & 1  & 0.47$\pm$0.28 & 2.1$\pm$0.2
   & 16 & 24$\pm$7      & 2.1$\pm$0.2
   & 7  & 5.7$\pm$1.6   & 0.93$\pm$0.17
   & 5  & 4.1$\pm$0.6   & 1.9$\pm$0.2\\
V & 0 & 0.21$\pm$0.12 & 1.9$\pm$0.2
  & 2 & 1.75$\pm$0.57 & 1.3$\pm$0.2
  & 1 & 0.66$\pm$0.37 & 0.70$\pm$0.14
  & 0 & 0.31$\pm$0.13 & 1.6$\pm$0.1\\
\end{tabular}
\end{ruledtabular}
\end{table*}

The presence of the third lepton in signal events can be used for
further separation from the background by requiring events to have a
third, isolated and well-measured track originating from the same
vertex as the two identified leptons. To maximize signal yield, no
additional lepton identification cuts are applied. The track (calorimeter)
isolation conditions for this third track have been
designed to be efficient for all lepton flavors, including
hadronic decays of $\tau$ leptons, by allowing for tracks (energy deposits)
inside an inner cone of $\dr<0.1$ ($\dr<0.2$). The distribution of the transverse
momentum \ptthree\ of the isolated track is shown in
Fig.~\ref{fig:selection2} for the \eel\ selection. 
Except for $W\!Z$ events, a third track in 
background events generally originates from the underlying event or jets, and
therefore tends to have very low transverse momentum.
$W\!Z$ events are suppressed by removing events where the third track
and one of the identified leptons have an invariant
mass~\mll\ consistent with the $Z$ boson mass~$M_Z$. For the
\lsmumu\ selection, backgrounds are low enough such that the
requirement of a third track is not needed. Instead, background from
$W\!Z\to\mu^{\pm}\nu\mu^{\pm}\mu^{\mp}$ is removed by vetoing events containing
opposite-sign muons with an invariant mass close to $M_Z$.
For the \mumul\ selection on the other hand, a significant amount of
multijet background remains; this is reduced by requiring that the
vectorial sum~$|\Sigma_{p_T}|$ of \met\ and muon transverse momenta balances the
transverse momentum of the third track.

Finally, a combined cut on the product of \met\ and \ptthree\ (\pttwo\
for \lsmumu) has been found to optimally reduce the remaining
background, which tends to have both low \met\ and low \ptthree.
The expected number of events for background and the reference signal 
defined above is summarized in Table~\ref{tab:numbers} at
various stages of the selection.  
After all cuts, the expected background is dominated by multijet background
(66\% and 53\% for the \mumul\ and \lsmumu\ selections, respectively) and
di-boson backgrounds (80\% and 88\% for \eel\ and \emul).

The estimate for expected number of background and signal events
depends on numerous measurements that each introduce a systematic
uncertainty: 
integrated luminosity (6.5\%), trigger efficiencies (1--2\%), 
lepton identification and reconstruction efficiencies (1--2\%), jet
energy scale calibration in signal ($<4$\%) and background events
(7--20\%), lepton and track momentum 
calibration (1\%), detector modeling (2\%), PDF uncertainties ($<4$\%),
and modeling of multijet background (4--40\%).
The uncertainties quoted in Table~\ref{tab:numbers} in addition contain 
the statistical uncertainty due to limited MC statistics, which 
is the dominant uncertainty for backgrounds from $W$ and $Z$ boson production.

As can be seen in Table~\ref{tab:numbers}, the numbers of events
observed in the data are in good agreement with the expectation from
standard model processes at all stages of the selection.
Combining all four selections, a total background of
2.93$\pm$0.54(stat)$\pm$0.57(syst) events is expected after all cuts, while 3
events are observed in the 
data. 

Since no evidence for associated production of charginos and
neutralinos is observed, an upper limit on the product of production
cross section and leptonic branching fraction~\sigbr\ is
extracted from this result. As mentioned above, information from the
four selections is combined using the modified frequentist
approach, taking into account correlated errors. The small
fraction of signal events that is selected by more than one
selection has been assigned to the selection with the largest
signal-to-background ratio and removed from all others.

\begin{figure}
\epsfig{file=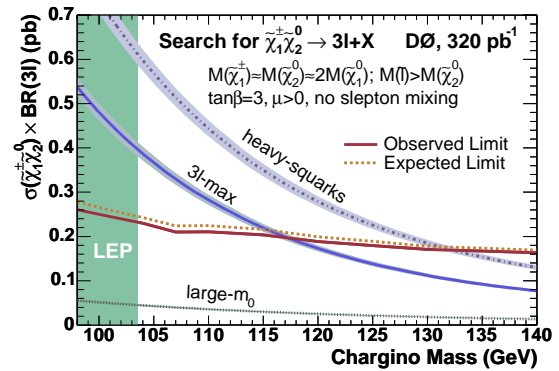,width=0.83\columnwidth,clip=true}
\caption{\label{fig:limit_mcha} Limit on \sigbr\ as a function of
chargino mass, in comparison with 
the expectation for several SUSY scenarios (see text). PDF and
renormalization/factorization scale
uncertainties are shown as shaded bands.}
\end{figure}
\begin{figure}
\epsfig{file=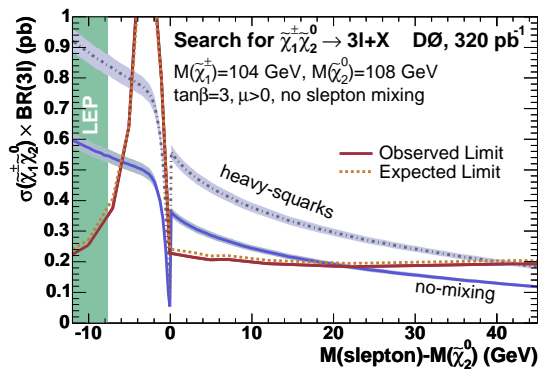,width=0.83\columnwidth,clip=true}
\caption{\label{fig:limit_dm} Limit on \sigbr\ as a function of the mass
difference between sleptons 
and $\chiz_2$, in comparison with the expectation for the MSSM (no mixing) and the
heavy-squarks scenario (see text). PDF and renormalization/factorization scale uncertainties are shown as shaded
bands. BR$(3\ell)$ drops sharply at \mslep$\lesssim$\mchitwo\ as the phase
space for two-body decays into real sleptons is minimal.
}
\end{figure}
The expected and observed limits are shown in
Figs.~\ref{fig:limit_mcha} and \ref{fig:limit_dm} as a function of
chargino mass and of the difference between chargino and slepton masses,
respectively. 
This result improves significantly the upper limit of about 1.5~pb set by
the D\O\ Run~I analysis \cite{run1_limits}.
Assuming the mSUGRA-inspired mass relation 
$\mcha\approx\mchitwo\approx 2\mchione$ as
well as degenerate slepton masses~\mslep\ (no slepton mixing), the limit on
\sigbr\ is a function 
of \mcha\ and \mslep, with a relatively small dependence on the other
SUSY parameters. This result can therefore be interpreted in more general SUSY
scenarios, as long as the above mass relations are satisfied and
$R$-parity is conserved.
The leptonic branching fraction of chargino and neutralino 
depends on the relative contribution from the slepton- and $W/Z$-exchange
graphs, which varies as a function of the slepton masses. $W/Z$ exchange is
dominant at large slepton masses, resulting in 
relatively small leptonic branching fractions ({\em large-$m_0$
scenario}). The leptonic branching fraction for three-body decays is
maximally enhanced for  
$\mslep\gtrsim\mchitwo$ ({\em 3$\,\ell$-max scenario}). Decays into
leptons can even be dominant if sleptons are light enough that
two-body decays are possible. 
In the latter case, one of the leptons from the neutralino decay can
have a very low transverse momentum if the mass 
difference between neutralino and sleptons is small.
In this region, only the \lsmumu\ selection remains efficient,
leading to a higher limit for $-6\lesssim\mslep-\mchitwo <0$~GeV
(see Fig.~\ref{fig:limit_dm}).
In addition, the $\cha_1\chiz_2$ production cross section depends on
the squark masses due to the negative interference with the $t$-channel
squark exchange. Relaxing scalar mass unification, the
cross section is maximal in the limit of large squark masses  ({\em
heavy-squarks scenario}). 
The NLO prediction~\cite{xsection} for \sigbr\ for these 
reference scenarios is shown in Figs.~\ref{fig:limit_mcha} and
\ref{fig:limit_dm}. The 
cross-section limit set in this analysis corresponds to a chargino mass
limit of 117~GeV (132~GeV) in the 3$\ell$-max (heavy-squarks) scenario,
which improves on the mass limit set by chargino searches at LEP.

In summary, no evidence for supersymmetry is observed in a
search for associated chargino and neutralino production in trilepton
events. Upper limits on the product of cross section and leptonic
branching fraction are set, which improve previous limits set
with the Run~I dataset. Chargino mass limits beyond the reach of LEP
chargino 
searches are derived for several SUSY reference scenarios with
enhanced leptonic branching fractions.

\input acknowledgement_paragraph_r2.tex   

%% file: acknowledgement_paragraph_r2.tex
%
We thank the staffs at Fermilab and collaborating institutions, 
and acknowledge support from the 
DOE and NSF (USA),
CEA and CNRS/IN2P3 (France),
FASI, Rosatom and RFBR (Russia),
CAPES, CNPq, FAPERJ, FAPESP and FUNDUNESP (Brazil),
DAE and DST (India),
Colciencias (Colombia),
CONACyT (Mexico),
KRF (Korea),
CONICET and UBACyT (Argentina),
FOM (The Netherlands),
PPARC (United Kingdom),
MSMT (Czech Republic),
CRC Program, CFI, NSERC and WestGrid Project (Canada),
BMBF and DFG (Germany),
SFI (Ireland),
A.P.~Sloan Foundation,
Research Corporation,
Texas Advanced Research Program,
Alexander von Humboldt Foundation,
and the Marie Curie Fellowships.